\documentclass[twocolumn, full paper]{jpsj3}
\usepackage{graphicx}
\usepackage{color}

\title{Anisotropic Uniaxial Pressure Response in UCoAl Studied by Nuclear Magnetic Resonance Measurement}

\author{
Kosuke \text{Karube}$^1$\thanks{E-mail: karube@scphys.kyoto-u.ac.jp}, 
Shunsaku Kitagawa$^{1,+}$,
Taisuke Hattori$^1$,  
Kenji Ishida$^1$\thanks{E-mail: kishida@scphys.kyoto-u.ac.jp}, \\
Noriaki Kimura$^{2,3}$, and 
Takemi Komatsubara$^{2,3}$}

\inst{$^1$Department of Physics, Graduate School of Science, Kyoto University, Kyoto 606-8502, Japan\\
$^2$Department of Physics, Graduate School of Science, Tohoku University, Sendai 980-8578, Japan\\
$^3$Center for Low Temperature Science, Tohoku University, Sendai 980-8578, Japan}

\abst{We have performed nuclear quadrupole resonance and nuclear magnetic resonance measurements on UCoAl with strong Ising-type anisotropy under $b$- and $c$-axes uniaxial pressure.
In the $b$-axis uniaxial pressure ($P_{\parallel b}$) measurement, we observed an increase in the metamagnetic transition field with increasing $P_{\parallel b}$.
In the $c$-axis uniaxial pressure ($P_{\parallel c}$) measurement, on the other hand, we observed a ferromagnetic transition in zero magnetic field along the $c$-axis above $P_{\parallel c}$ = 0.08 GPa.
The anomaly of the nuclear spin-lattice relaxation rate divided by the temperature $\left[ (T_1 T)^{-1} \right]$ at $T$ = 20 K is suppressed by $P_{\parallel b}$ and slightly enhanced by $P_{\parallel c}$.
The anisotropic uniaxial pressure response indicates that uniaxial pressure is a good parameter for tuning the Ising magnetism in UCoAl.}

\kword{UCoAl, uniaxial pressure, NQR, NMR}

\begin{document}
\maketitle

\section{Introduction}
Itinerant ferromagnetic (FM) compounds have attracted much attention because some of them exhibit exotic ordered states, for example, unconventional superconductivity in UGe$_2$\cite{Saxena}, a skyrmion lattice ordered state in MnSi\cite{Muhlbauer}, and a nematic state in Sr$_3$Ru$_2$O$_7$\cite{Grigera}.
These FM compounds seem to follow the temperature ($T$) $-$ magnetic field ($H$) $-$ pressure ($P$) 3-dimensional (3D) universal phase diagram\cite{Taufour,Pfleiderer,Wu}.
In this phase diagram, the second-order FM transition line in zero magnetic field bifurcates into finite magnetic fields at the tricritical point (TCP) with the formation of first-order transition planes and finally terminates at the quantum critical endpoint (QCEP) at zero temperature. 

\begin{figure}[tbp]
\begin{center}
\includegraphics[width=7cm]{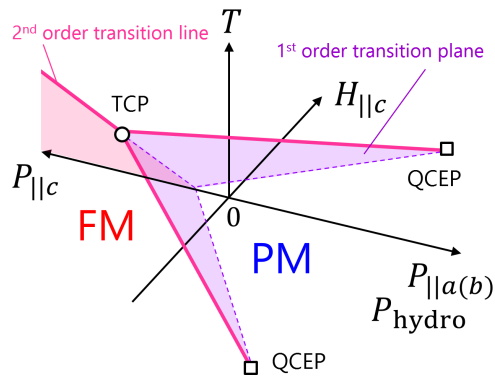}
\end{center}
\caption{(Color online) 3D schematic phase diagram for UCoAl with temperature ($T$) $-$ magnetic field along the $c$-axis ($H_{\parallel c}$) $-$ hydrostatic pressure ($P_\mathrm{hydro}$) and uniaxial pressure along the $a(b)$- and $c$-axes ($P_{\parallel a(b)}$,$P_{\parallel c}$). The pink line shows the second-order transition line and the purple plane shows the first-order transition plane. The TCP and QCEP are denoted as a circle and square, respectively.}
\label{fig_THP_PD_UCoAl}
\end{figure}

UCoAl, as reported here, is believed to be an itinerant FM compound and its 3D phase diagram is depicted in Fig.~\ref{fig_THP_PD_UCoAl}.
UCoAl possesses the hexagonal ZrNiAl-type (space group: $P\overline{6}2m$, No. 189) crystal structure with alternately stacked U-Co(1) and Co(2)-Al layers; here we define the [0$\overline{1}$10], [1000], and [0001] directions in the hexagonal structure as the $a$-, $b$-, and $c$-axes, respectively.
At ambient pressure, its ground state is paramagnetic (PM) with strong Ising-type anisotropy (the easy magnetization axis is the $c$-axis), 
but it undergoes a first-order metamagnetic transition (field-induced FM transition) under a relatively small magnetic field of $\mu_0 H_{\parallel c}$ $\sim$ 0.6 T only applied along the $c$-axis. 
The first-order metamagnetic transition line in the $T$ $-$ $H_{\parallel c}$ phase diagram terminates at the critical endpoint (CEP) at 12 K and changes to a crossover.
Nuclear magnetic resonance (NMR) studies revealed that magnetic fluctuations are also strongly Ising-type along the $c$-axis and diverge at the CEP\cite{Nohara,Karube}.

Hydrostatic pressure measurements on UCoAl have been intensively carried out\cite{Mushnikov,Aoki}.
Recently, Aoki \textit{et al}. quantitatively reported that under hydrostatic pressure the CEP moves to a lower temperature and a higher field and finally reaches the QCEP at ($T$, $\mu_0$$H_{\parallel c}$, $P_\mathrm{hydro}$)$_\mathrm{QCEP}$ $\sim$ (0 K, 7 T, 1.5 GPa)\cite{Aoki}.
Uniaxial pressure measurements on UCoAl are rare in comparison with hydrostatic pressure measurements because of experimental difficulties.
Saha \textit{et al}. reported that $a(b)$-axis uniaxial pressure has the same effect as hydrostatic pressure\cite{Saha}. 
On the other hand, Ishii \textit{et al}. reported that the $c$-axis uniaxial pressure induces the FM transition in zero magnetic field\cite{Ishii}, which implies that $c$-axis uniaxial pressure is the opposite tuning parameter to hydrostatic pressure and $a(b)$-axis uniaxial pressure in the 3D phase diagram.
Although UCoAl is one of the suitable compounds for figuring out the riddle of the universal 3D phase digram for itinerant FM compounds, there have been insufficient uniaxial pressure studies on UCoAl reported so far. Here we report static and dynamic magnetic properties of UCoAl under $b$-axis and $c$-axis uniaxial pressure studied by nuclear quadrupole resonance (NQR) and NMR measurements.
\section{Experimental Procedure}
\begin{figure}[tbp]
\begin{center}
\includegraphics[width=9cm]{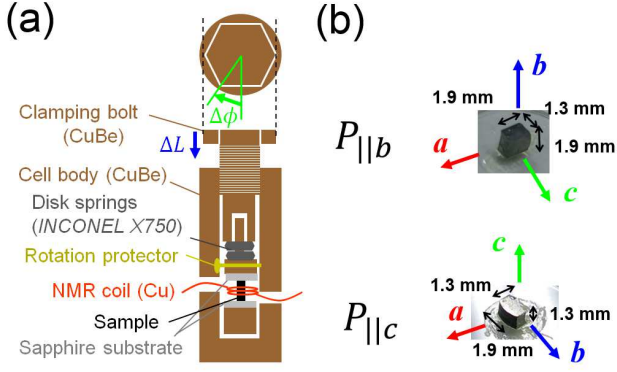}
\end{center}
\caption{(Color online) (a) Schematic figure of the pressure cell. $\Delta L$ and $\Delta \phi$ represent the travel distance and the rotation angle of the clamping bolt, respectively. (b) Pictures of the single-crystal sample used for each pressure direction.}
\label{fig_cell_sample}
\end{figure}
For the uniaxial pressure measurements, we constructed a pressure clamp cell from a hardened non-magnetic CuBe alloy as schematically shown in Fig.~\ref{fig_cell_sample}(a). 
An uniaxial force was created by tightening a pressure clamping bolt and transferred to the sample by non-magnetic disk springs (INCONEL X750). The sample was sandwiched between two sapphire substrates, and a Cu coil to detect NQR and NMR signals was wound around the sample parallel to the pressure direction.
In order to avoid damaging the sample by rotation of the inside jig, the jig was pinned by a screw (rotation protector) whose head was stopped at the window in the cell body.
The magnitude of uniaxial pressure along the $i$-axis $P_{\parallel i}$ was determined by Hooke's law at room temperature as $P_{\parallel i} = (k \cdot \Delta L)/S_i$, where $k$ = 3678.5 N/mm is the spring constant and $S_i$ is the surface area of the $i$-plane ($S_b$ = $S_c$ = 2.47 mm$^2$ in the present UCoAl samples).
The shrinkage of the spring $\Delta L$ (on the order of 0.1 mm) was identified as the travel distance of the clamping bolt attached to the inside jig and was estimated by measuring the rotation angle of the clamping bolt $\Delta \phi$ using the relation $\Delta L = p\cdot (\Delta \phi/360^\circ)$, where $p$ = 1.75 mm is the pitch (= lead) of the clamping bolt.

We also investigated the temperature dependence of pressure by measuring the resistance of a strain gauge attached to Cu as a reference material ($S$ = 18.8 mm$^2$) pressurized in the present pressure cell. 
The pressure estimated by the strain gauge at temperature $T$ is given by the following equation: 
\begin{eqnarray}
P (T) = E_\mathrm{Cu} \cdot \frac{1}{k_\mathrm{SG}}\frac{R_0 (T)-R (T)}{R_0 (T)}, 
\end{eqnarray}
where $E_\mathrm{Cu}$ = 120 GPa is the Young's modulus of Cu, $k_\mathrm{SG}$ = 2 is the transformation ratio from the strain to the resistance in the  strain gauge, and $R_0 (T)$ and $R (T)$ are the resistance of the strain gauge at ambient pressure and under uniaxial pressure, respectively.
The relative change in $P (T)$ from room temperature (R.T.) is in the range of $0 < [P (T)-P (\mathrm{R.T.})]/P (\mathrm{R.T.}) < 0.1$ down to $T$ = 4.2 K, suggesting that the temperature dependence of uniaxial pressure in the present measurement is relatively small, although the pressure is not exactly temperature independent due to effects such as thermal expansion and the temperature dependence of the spring constant.

A single-crystal UCoAl sample was synthesized by the Czochralski pulling method in a tetra-arc furnace.
For the present uniaxial pressure study, the sample was cut to a rectangular shape, as shown in Fig.~\ref{fig_cell_sample}(b), with dimensions of 
1.9 ($a$-axis) $\times$ 1.9 ($b$-axis) $\times$ 1.3 ($c$-axis) mm$^3$ for uniaxial pressure along the $b$-axis and 
1.9 ($a$-axis) $\times$ 1.3 ($b$-axis) $\times$ 1.3 ($c$-axis) mm$^3$ for uniaxial pressure along the $c$-axis.
The surfaces of the $b$- and $c$-planes were polished to ensure homogeneous pressure.

Using the above pressure cell and UCoAl samples, we performed $^{27}$Al-NMR measurements controlling three parameters: temperature ($T$), magnetic field along the $c$-axis ($H_{\parallel c}$), and uniaxial pressure along the $b$- and $c$-axes ($P_{\parallel b,c}$). $H_{\parallel c}$ was tuned by the angle $\theta$ between the magnetic field $H$ and the crystal $c$-axis using the relation $H_{\parallel c} = H\cos\theta$ since the physical properties are insensitive to the magnetic field perpendicular to the $c$-axis. In all measurements, the NMR radio frequency was fixed at $f$ = 17.5 MHz [$\mu_0 H$($^{27}K$ = 0) = 1.577 T].
\begin{figure}[tbp]
\begin{center}
\includegraphics[width=8cm]{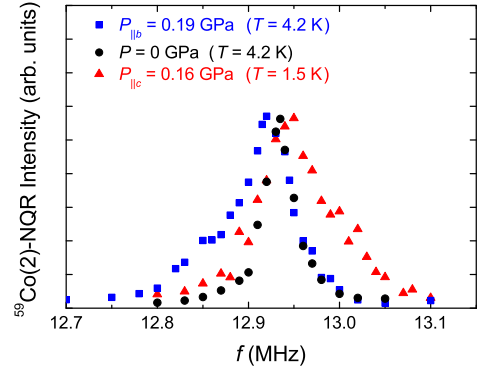}
\end{center}
\caption{(Color online) Zero-magnetic-field $^{59}$Co(2)-NQR spectra of the $\nu_3$ peak at $P$ = 0 GPa (circles), $P_{\parallel b}$ = 0.19 GPa (squares), and  $P_{\parallel c}$ = 0.16 GPa (triangles).}
\label{fig_NQR}
\end{figure}
\section{Results }
Before showing the $^{27}$Al-NMR results, we show $^{59}$Co(2)-NQR results obtained in zero magnetic field. 
The NQR reflects the interaction between the quadrupole moments of the nuclear spin $I$ and the electric field gradient (EFG) in the crystal, and its Hamiltonian is described as follows:
\begin{eqnarray}
\mathcal{H}_\mathrm{Q}=\frac{\hbar \nu_{zz}}{6}\left\{ (3I_z^2 - I^2)+\frac{\eta}{2}(I_{+}^2+I_{-}^2) \right\},
\end{eqnarray}
where $\nu_{zz}$ $(\propto V_{zz})$ is the quadrupole frequency along the EFG principal axis and $\eta$ $(= \left| V_{xx}-V_{yy} \right| / V_{zz})$ is the asymmetry parameter with respect to the EFG principal axis; here $V_{ij}$ $\equiv$ $\partial^2 V/\partial x_i \partial x_j$ ($V$ is the electrostatic potential; $x_i, x_j = x, y, z$) is the EFG tensor.
In UCoAl, the EFG principal axis is parallel to the $c$-axis, and for $^{59}$Co(2) nuclei $\nu_{zz}$ = 4.312 MHz and $\eta$ = 0, which yield three NQR signals at $\nu_1$ = 4.312 MHz, $\nu_2$ = 8.623 MHz, and $\nu_3$ = 12.935 MHz\cite{Nohara, Karube, Iwamoto}. 
In the present study, we tracked the $\nu_3$ signal at $T$ = 4.2 and 1.5 K under $P$ = 0 GPa, $P_{\parallel b}$ = 0.19 GPa, and  $P_{\parallel c}$ = 0.16 GPa.
As shown in Fig.~\ref{fig_NQR}, the sharp $\nu_3$ spectrum at ambient pressure was broadened toward a lower frequency (smaller $V_{zz}$) by $P_{\parallel b}$, on the other hand, it was broadened toward a higher frequency (larger $V_{zz}$) by $P_{\parallel c}$. This anisotropic response of the NQR spectra to uniaxial pressure suggests that the external uniaxial pressures create microscopic uniaxial strains along the pressure directions.
However, Fig.~\ref{fig_NQR} also indicates that uniaxial pressures possess inhomogeneity ranging from zero to a finite value, because the NQR spectra under the uniaxial pressures were broadened on both sides from the frequencies observed in the original spectra at ambient pressure.
We consider that the origin of the inhomogeneous pressure is the cubic shape of the present samples, i.e., uniaxial strain is maximized around the central region of the sample and becomes smaller toward both pressurized surfaces.
Thus, much longer samples along the uniaxial pressure directions are desired to avoid pressure inhomogeneity. 
\begin{figure}[tbp]
\begin{center}
\includegraphics[width=8cm]{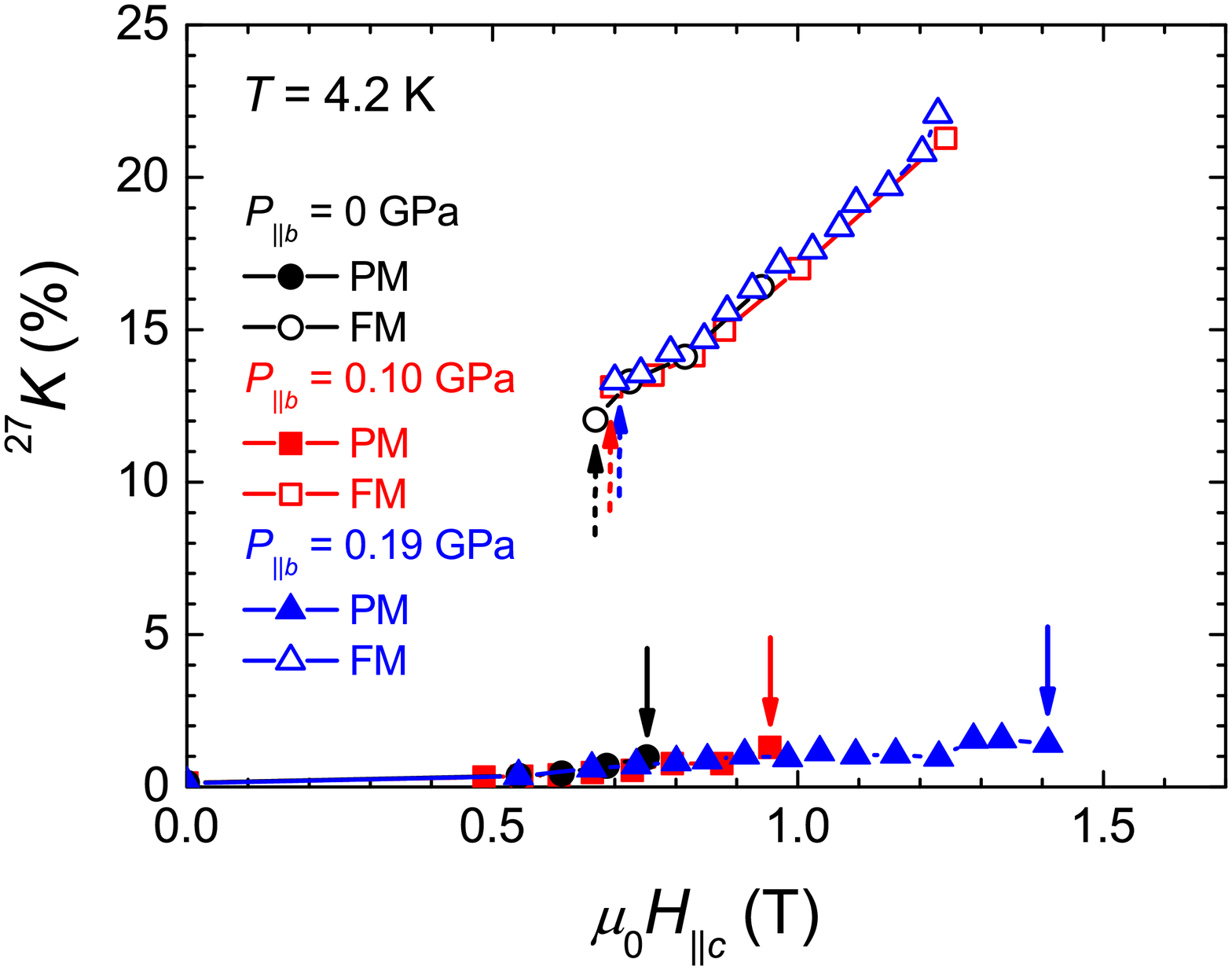}
\end{center}
\caption{(Color online) $H_{\parallel c}$ dependence of the $^{27}$Al-NMR Knight shift $^{27}K$ at $T$ = 4.2 K for $P_{\parallel b}$ = 0 GPa (circles), 0.10 GPa (squares), and 0.19 GPa (triangles). $^{27}K$ for the PM and FM spectra are denoted as closed and open symbols, respectively. The solid and dashed arrows represent $H_{\parallel c}^\mathrm{PM dis}$ ($H_{\parallel c}$ where the PM spectra disappear) and $H_{\parallel c}^\mathrm{FM app}$ ($H_{\parallel c}$ where the FM spectra appear), respectively.}
\label{fig_KvsHc_Pb}
\end{figure}

\begin{figure}[tbp]
\begin{center}
\includegraphics[width=8cm]{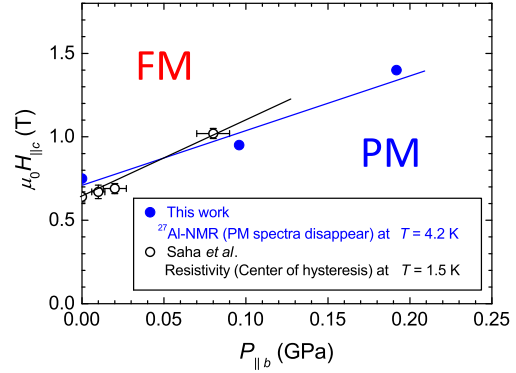}
\end{center}
\caption{(Color online) $H_{\parallel c}$ $-$ $P_{\parallel b}$ phase diagram at $T$ = 4.2 K obtained by the present NMR measurement (by plotting $H_{\parallel c}^\mathrm{PM dis}$) (closed symbols) with the previous resistivity measurement by Saha \textit{et al}.\cite{Saha} (open symbols).}
\label{fig_HcvsPb}
\end{figure}

We measured the field-swept $^{27}$Al-NMR spectra under various $\theta$ ($H_{\parallel c} = H\cos\theta$) for $P_{\parallel b}$ = 0, 0.10, and 0.19 GPa at $T$ = 4.2 K and estimated Knight shift $^{27}K$ against $H_{\parallel c}$ as shown in Fig.~\ref{fig_KvsHc_Pb}.
At ambient pressure, PM spectra disappear at $\mu_0 H_{\parallel c}$ $\sim$ 0.7 T and FM spectra appear in the separated field region, which is a typical feature of a first-order metamagnetic transition as reported in our previous NMR study\cite{Karube}. With increasing $P_{\parallel b}$, $H_{\parallel c}^\mathrm{PM dis}$ ($H_{\parallel c}$ where the PM spectra disappear, denoted as the solid arrows in Fig.~\ref{fig_KvsHc_Pb}) increases, which indicates that the metamagnetic transition field increases with increasing $P_{\parallel b}$. However, $H_{\parallel c}^\mathrm{FM app}$ ($H_{\parallel c}$ where FM spectra appear, denoted as the dashed arrows in Fig.~\ref{fig_KvsHc_Pb}) is insensitive to $P_{\parallel b}$. This is because of the inhomogeneity of the uniaxial pressure as shown by the above NQR results. Therefore, in the present measurement, $H_{\parallel c}^\mathrm{PM dis}$ is the marker of the metamagnetic transition field and was plotted in the $H_{\parallel c}$ $-$ $P_{\parallel b}$ phase diagram along with the previous result by Saha \textit{et al}.\cite{Saha} as shown in Fig.~\ref{fig_HcvsPb}.

\begin{figure}[tbp]
\begin{center}
\includegraphics[width=8cm]{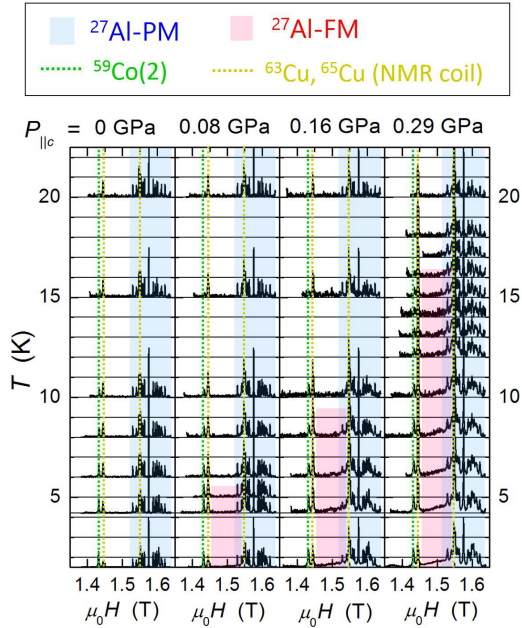}
\end{center}
\caption{(Color online) Field-swept NMR spectra for several $T$ and $P_{\parallel c}$ at $H_{\parallel c}$ = 0. The blue and red shaded spectra represent the $^{27}$Al-PM and FM spectra, respectively. The green dotted line represents one of the $^{59}$Co(2) satellite peaks. The yellow dotted lines represent $^{63}$Cu ($\mu_0 H$ $\sim$ 1.55 T) and $^{65}$Cu ($\mu_0 H$ $\sim$ 1.44 T) signals arising from the NMR coil. All spectra were normalized by the intensity of the central peak of the $^{27}$Al-PM spectrum.}
\label{fig_SpvsT_Pc}
\end{figure}

\begin{figure}[tbp]
\begin{center}
\includegraphics[width=8cm]{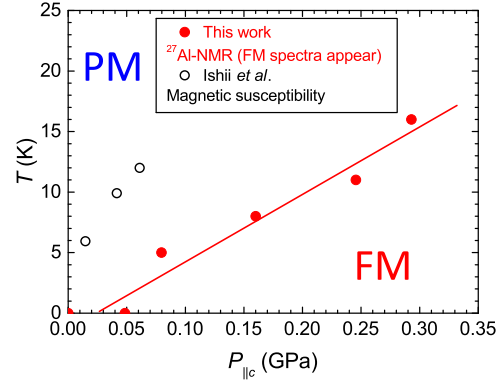}
\end{center}
\caption{(Color online) $T$ $-$ $P_{\parallel c}$ phase diagram at $H_{\parallel c}$ = 0 obtained by the present NMR measurement (by plotting $T^\mathrm{FM app}$) (closed symbols) with the previous magnetic susceptibility measurement by Ishii \textit{et al}.\cite{Ishii} (open symbols).}
\label{fig_TvsPc}
\end{figure}

\begin{figure}[tbp]
\begin{center}
\includegraphics[width=9cm]{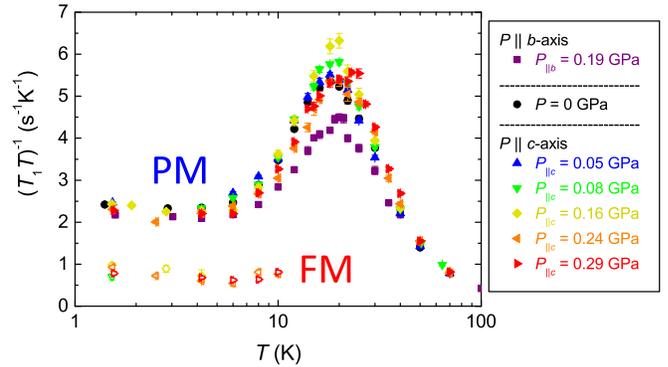}
\end{center}
\caption{(Color online) Temperature dependence of $(T_1T)^{-1}$ for several uniaxial pressures at $H_{\parallel c}$ = 0. The closed and open symbols represent $(T_1T)^{-1}$ in the $^{27}$Al-PM and FM spectra, respectively. }
\label{fig_NMRinvT1TvsT_Pbc}
\end{figure}

We measured the field-swept $^{27}$Al-NMR spectra at various $T$ and $P_{\parallel c}$ as shown in Fig. \ref{fig_SpvsT_Pc}.
The magnetic field was applied along the $a$-axis ($H_{\parallel c}$ = 0). 
All spectra were normalized by the intensity of the central peak of the $^{27}$Al-PM spectrum at $\mu_0 H$ $\sim$ 1.575 T ($^{27}K$ $\sim$ 0.1\%).
With increasing $P_{\parallel c}$, the peak intensity of the $^{27}$Al-PM spectra decreases and broad spectra appear around $\mu_0 H$ $\sim$ 1.5 T ($^{27}K$ $\sim$ 5\%) above $P_{\parallel c}$ = 0.08 GPa. 
As shown in Fig.~\ref{fig_NMRinvT1TvsT_Pbc}, the nuclear spin-lattice relaxation time $T_1$ in the broad spectra is approximately three times longer than that in the sharp $^{27}$Al-PM spectra. 
The longer $T_1$ is evidence of the broad spectra originating from the FM components.
With increasing $P_{\parallel c}$, $T^\mathrm{FM app}$ (the temperature where the FM spectra appear around $\mu_0 H$ $\sim$ 1.5 T) increases.
However, the PM spectra remain down to $T$ = 1.5 K and up to $P_{\parallel c}$ = 0.29 GPa and the intensity ratio of the FM spectra relative to the PM spectra is small, $\sim$ 13\% at ($T$, $P_{\parallel c}$) = (1.5 K, 0.29 GPa).
This is also because of the inhomogeneity of pressure as shown by the above NQR results; thus, it is difficult to distinguish whether the FM transitions are first-order or second-order from the NMR spectra. 
Therefore, in this measurement, $T^\mathrm{FM app}$ is the marker of the FM transition temperature and was plotted in the $T$ $-$ $P_{\parallel c}$ phase diagram along with the previous result by Ishii \textit{et al}.\cite{Ishii} as shown in Fig.~\ref{fig_TvsPc}. 
The FM transition line is not consistent with the previous study.
This is because of the dependences of samples or the determination of FM transition temperatures.

We emphasize that $P_{\parallel b,c}$ (the horizontal axes of Figs.~\ref{fig_HcvsPb} and \ref{fig_TvsPc}) are calculated with the assumption of the ideal homogeneous uniaxial pressure but they are actually the average values of inhomogeneous uniaxial pressure. On the other hand, $H_{\parallel c}^\mathrm{PM dis}$ and $T^\mathrm{FM app}$ (the vertical axes of Figs.~\ref{fig_HcvsPb} and \ref{fig_TvsPc}) are evaluated with the maximum values of the inhomogeneous pressure. 
This evaluation with inhomogeneity taken into account is impossible for macroscopic measurements but possible for the microscopic probe in the present NMR technique.


\begin{figure}[tbp]
\begin{center}
\includegraphics[width=9cm]{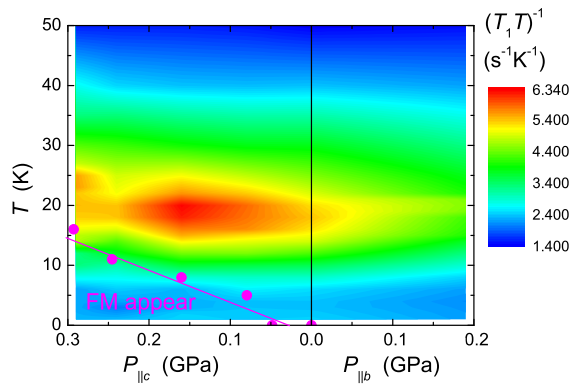}
\end{center}
\caption{(Color online) Contour plot of $(T_1T)^{-1}$ of $^{27}$Al-PM spectra in the $T$ $-$ $P_{\parallel b,c}$ ($H_{\parallel c}$ = 0) phase diagram.
FM transition points $T^\mathrm{FM app}$ obtained in the present NMR study are denoted by circles.}
\label{fig_contourplot}
\end{figure}

The nuclear spin-lattice relaxation time $T_1$ was measured at the central peak of the $^{27}$Al-NMR spectra (at $\mu_0 H$ = 1.575 T for the PM spectra and at $\mu_0 H$ = 1.5 T for the FM spectra), corresponding to the transition between the nuclear spin states $I$ = 1/2 and -1/2.
For the measurement of $T_1$, the nuclear magnetization after the saturation pulses was fitted with the theoretical recovery function.
In this measurement, the magnetic field was applied along the $a$-axis ($H_{\parallel c}$ = 0).
$(T_1T)^{-1}$ is used to probe hyperfine-field fluctuations perpendicular to the applied fields. 
In addition, our previous NMR study\cite{Karube} revealed that UCoAl has strong Ising-type magnetic fluctuations along the $c$-axis that are independent of the magnetic field along the $a$-axis.
Thus, $(T_1T)^{-1}$ in this measurement represents the magnetic fluctuations along the $c$-axis [$(T_1T)^{-1}_{H \parallel a} = S_{\parallel b} + S_{\parallel c} \sim S_{\parallel c}$ (since $S_{\parallel c}$ $\gg$ $S_{\parallel b}$), where $S_{\parallel i}$ is the magnetic fluctuations along the $i$-axis at the $^{27}$Al nuclei sites]. 
Figure \ref{fig_NMRinvT1TvsT_Pbc} shows the temperature dependence of $(T_1T)^{-1}$ at ambient pressure and several $P_{\parallel b}$ and $P_{\parallel c}$. 
In addition, we constructed a contour plot of $(T_1T)^{-1}$ for the PM spectra in the $T$ $-$ $P_{\parallel b,c}$ phase diagram as shown in Fig.~\ref{fig_contourplot}.
We show the presence of a broad peak of $(T_1T)^{-1}$ around 20 K in the entire pressure range.
This characteristic temperature $T_\mathrm{max}$, where magnetic fluctuations [often seen in magnetic susceptibility $\chi$ and $(T_1T)^{-1}$] exhibit a broad maximum, is typically observed in itinerant metamagnetic compounds, for example, $T_\mathrm{max}$ $\sim$ 260 K in YCo$_2$\cite{Burzo}, 80 K in Co(S$_{1-x}$Se$_x$)$_2$\cite{Goto}, and 20 K in UCoAl\cite{Nohara, Karube, Mushnikov} in the present case.
Note that the peak intensity of $(T_1T)^{-1}$ at $T_\mathrm{max}$ $\sim$ 20 K is suppressed with increasing $P_{\parallel b}$ but slightly enhanced with increasing $P_{\parallel c}$ and maximized around $P_{\parallel c}$ = 0.16 GPa.
\section{Discussion}
We reproduced the results of the previous uniaxial pressure studies and clarified the anisotropic response for uniaxial pressure directions from the microscopic viewpoint by the NMR measurements. 
This is strong evidence that the uniaxial pressure is a good tuning parameter in the 3D phase diagram for UCoAl.
The anisotropic response to the uniaxial pressure suggests that the anisotropic hybridization of U-5$f$ electrons in the hexagonal crystal structure plays an important role in the magnetism of UCoAl.
As discussed in the previous reports\cite{Saha, Ishii}, UCoAl has strong intraplane 5$f$-3$d$ (U-Co(1)) hybridization in the $ab$-plane, whereas the interplane  5$f$-3$d$ (U-Co(2)) and  5$f$-3$p$ (U-Al) hybridizations along the $c$-axis are much weaker than the intraplane hybridization. Thus, the magnetic properties are determined by the intraplane hybridization. 
As represented by the Hill criterion\cite{Hill} (the nearest U-U distance in the $ab$-plane in UCoAl, $d_\mathrm{U-U}$ $\sim$ 3.46 $\mathrm{\mathring{A}}$\cite{Javorsky}, is in the range of the Hill limit, 3.4-3.6 $\mathrm{\mathring{A}}$), 
the further enhancement of the intraplane hybridization by applying uniaxial pressure along the $ab$-plane produces the itinerant behavior with 5$f$-bands and suppresses the FM ordered state. On the other hand, the weakening of the intraplane hybridization by applying uniaxial pressure along the $c$-axis produces localized behavior with 5$f$-moments and enhances the FM ordered state.

We interpreted the results of the nuclear relaxation rates as follows. 
In the $T$ $-$ $P_{\parallel b,c}$ plane at $H_{\parallel c}$ = 0, magnetic fluctuations are expected to be maximized at the TCP; thus, 
the results suggest that the TCP exists at approximately $P_{\parallel c}$ $\sim$  0.16 GPa.
However, the displayed $(T_1T)^{-1}$ is the average of those of various pressure values because of the inhomogeneity of pressure; 
thus, further uniaxial pressure measurements without the inhomogeneity of pressure are desired to quantitatively discuss the criticality of the TCP.
\section{Conclusion}
We have performed $^{27}$Al-NMR measurements on UCoAl under $b$- and $c$-axes uniaxial pressure with a homemade uniaxial pressure clamp cell. Both the static and dynamic magnetic properties exhibit the opposite responses for different uniaxial pressure directions.
The $b$-axis uniaxial pressure increases the metamagnetic transition field and magnetic fluctuations are suppressed.
The $c$-axis uniaxial pressure induces the FM transition in zero magnetic field along the $c$-axis and magnetic fluctuations are slightly enhanced and maximized at 0.16 GPa.
These results indicate that the uniaxial pressure is a good parameter for tuning the strongly anisotropic magnetism in UCoAl.
The maximum of the magnetic fluctuations under the $c$-axis uniaxial pressure suggests the existence of a TCP on the FM transition line.
\section*{Acknowledgments}
We are grateful to D. Aoki, H. Kotegawa, A. V. Andreev, S. Yonezawa, and Y. Maeno for fruitful discussions. 
This work is supported by Grants-in-Aid for Scientific Research on Innovative Areas ``Heavy Electrons" (No.~20102006) from MEXT, for the GCOE Program ``The Next Generation of Physics, Spun from Universality and Emergence" from MEXT, and for Scientific Research from JSPS. 
One of the authors (KK) was financially supported by a JSPS Research Fellowship.

\vspace{1cm}
\noindent
+Present address: Department of Physics, Faculty of Science, Okayama University, Okayama 700-8530, Japan

\end{document}